\documentclass[
reprint,
showpacs,
superscriptaddress,
 amsmath,amssymb,
 aip,
 pop,
]{revtex4-1}

\usepackage{graphicx}
\usepackage{dcolumn}
\usepackage{color}
\usepackage{bm}
\usepackage[normalem]{ulem}

\begin{document}

\preprint{APS/123-QED}

\title{Synchrotron radiation, pair production and longitudinal electron motion during 10-100PW laser solid interactions}
\author{C. S. Brady} \affiliation{Physics Department, University of Warwick, Coventry, CV4 7AL, UK}
\author{C. P. Ridgers}
\affiliation{University of Oxford} \affiliation{University of York}
\author{T. D. Arber} \affiliation{Physics Department, University of Warwick, Coventry, CV4 7AL, UK}
\author{A. R. Bell}
\affiliation{University of Oxford}
\date{\today}
\begin{abstract}
At laser intensities above $10^{23} \mathrm{W/cm^2}$ the interaction of a laser with a plasma is qualitatively different to the interactions at lower intensities. In this intensity regime solid targets start to become relativistically underdense, gamma-ray production by synchrotron emission starts to become an important feature of the dynamics and, at even higher intensities, electron-positron pair production by the non-linear Breit-Wheeler process starts to occur. Previous work in this intensity regime has considered ion acceleration \cite{Chen2011},\cite{Salamin2008}, identified different mechanisms for the underlying plasma physics of laser generation of gamma-rays  \cite{Ridgers2011}, \cite{Brady2012}, \cite{Brady2013} considered the effect of target parameters on gamma-ray generation \cite{Nakamura2012} and considered the creation of solid density positronium plasma \cite{Ridgers2011}. However a complete linked understanding of the important new physics of this regime is still lacking. In this paper, an analysis is presented of the effects of target density, laser intensity, target preplasma properties and other parameters on the conversion efficiency, spectrum and angular distribution of gamma-rays by synchrotron emission. An analysis of the importance of Breit-Wheeler pair production is also presented. Target electron densities between $10^{22} \mathrm{cm^{-3}}$ and $5 \times 10^{24} \mathrm{cm^{-3}}$ and laser intensities covering the range between $10^{21} \mathrm{W/cm^2}$ (available with current generation laser facilities) and $10^{24} \mathrm{W/cm^2}$ (upper intensity range expected from the ELI facility) are considered. It is found that peak efficiency of conversion of laser energy into gamma-ray energy is achieved when the target density is 0.1 times the relativistically corrected critical density and that higher efficiency is obtained at higher laser intensity. Target front surface preplasmas of sufficient length are found to increase the efficiency of gamma-ray production compared with striking overdense solid targets directly in a fashion comparable to that in Nakamura et al. \cite{Nakamura2012} by ensuring that the laser pulse interacts with a plasma of the optimum density. However maximum laser conversion to gamma-rays is achieved by striking a uniform target of the optimum density or a preplasma of sufficient length that the preplasma is nearly a uniform slab. It is found that the efficiency of Breit-Wheeler pair production is not related to either relativistic transparency or efficiency of gamma-ray production but is maximized by striking a high density target with the most intense laser possible. A qualitative model for this behaviour is presented. An analysis of the behavior of the longitundal motion of electrons as target density and laser intensity change is presented and it is found that there are two distinct regimes separated by the ratio of the relativistically corrected plasma frequency to the laser frequency.  It is found that the transition between the two regimes is related to the structure of the electron phase space at the head of the laser.
\end{abstract}
\maketitle
\section{Introduction}
The maximum achievable laser intensity has climbed steadily since the first application of chirped pulse amplification (CPA) to high power laser systems and is expected to reach $10^{23} \mathrm{W/cm^2}$ within the next few years. Many possible applications have been proposed for lasers of this intensity, ranging from some fast ignition schemes \cite{Mourou2007} to ion acceleration to energies required for medical applications \cite{Dromey2006}. In order to realize these possibilities accurate modelling of the laser-plasma interactions is necessary, but this is complicated by the fact that at these intensities additional physical processes beyond those described by the Vlasov-Maxwell system of equations starts to become important. Emission of short wavelength radiation by non-linear Compton scattering from highly relativistic electrons starts to become important and the requirement that this emission be modeled in a quantum electrodynamically correct way leads to this new regime sometimes being called the QED-plasma regime. Some possible applications of lasers in this intensity regime take advantage of these processes to either act as a high brightness gamma-ray source \cite{Brady2012} or even as a positronium factory \cite{Ridgers2011}. At an intensity of $10^{23} \mathrm{W/cm^2}$ the relativistic gamma factor of an electron quivering in the field of a 1.06 micron laser can be as high as 200 leading to a relativistically corrected critical density of $2\times 10^{23} \mathrm{cm^{-3}}$. This is higher than the electron number density of lithium ($1.3\times 10^{23} \mathrm{cm^{-3}}$), about two thirds that of solid polystyrene ($3.6\times 10^{23}\mathrm{cm^{-3}}$) and about a quarter that of solid aluminium ($7.7\times 10^{23}\mathrm{cm^{-3}}$) meaning that at and around this intensity solid targets start to become relativistically transparent. While the lasers needed to access this regime have not yet been built, there is still theoretical interest in describing this new regime. Previous work has included the effects of photon emission on ion acceleration \cite{Chen2011}, methods of generating dense positronium plasmas \cite{Ridgers2011} and the general theory underlying the controlling parameters for photon emission \cite{Nakamura2012}, \cite{Brady2012}. This previous work has demonstrated an association between strong photon emission and low-density plasma (in preplasmas for Nakamura et al. \cite{Nakamura2012} and directly in low-density plasma for Brady et al. \cite{Brady2012}), particularly plasma that is relativistically underdense. What is missing is a general overview of how QED-plasmas behave as a general guide to where in parameter space should be considered for different applications. This paper attempts to provide such an overview.\\

This paper is split into four sections. The first covers the previous work on the mechanisms responsible for gamma-ray generation. The second deals with the parameters which affect the creation of gamma-rays by synchrotron emission. The third details the controlling parameters for the production of electron-positron pairs by the non-linear Breit-Wheeler process. Since synchrotron emission and pair production are strongly affected by the the ratio of the target density to the relativistically corrected critical density the final section presents an overview of how the longitudinal motion of the electrons controls the two QED processes. The results in all of these sections are based on 1D and 2D particle in cell simulations using the QED-PIC code EPOCH \cite{Brady2011}, \cite{Bennett2013}.

\section{gamma-ray emission models}

\begin{figure} 
\includegraphics[scale=0.5]{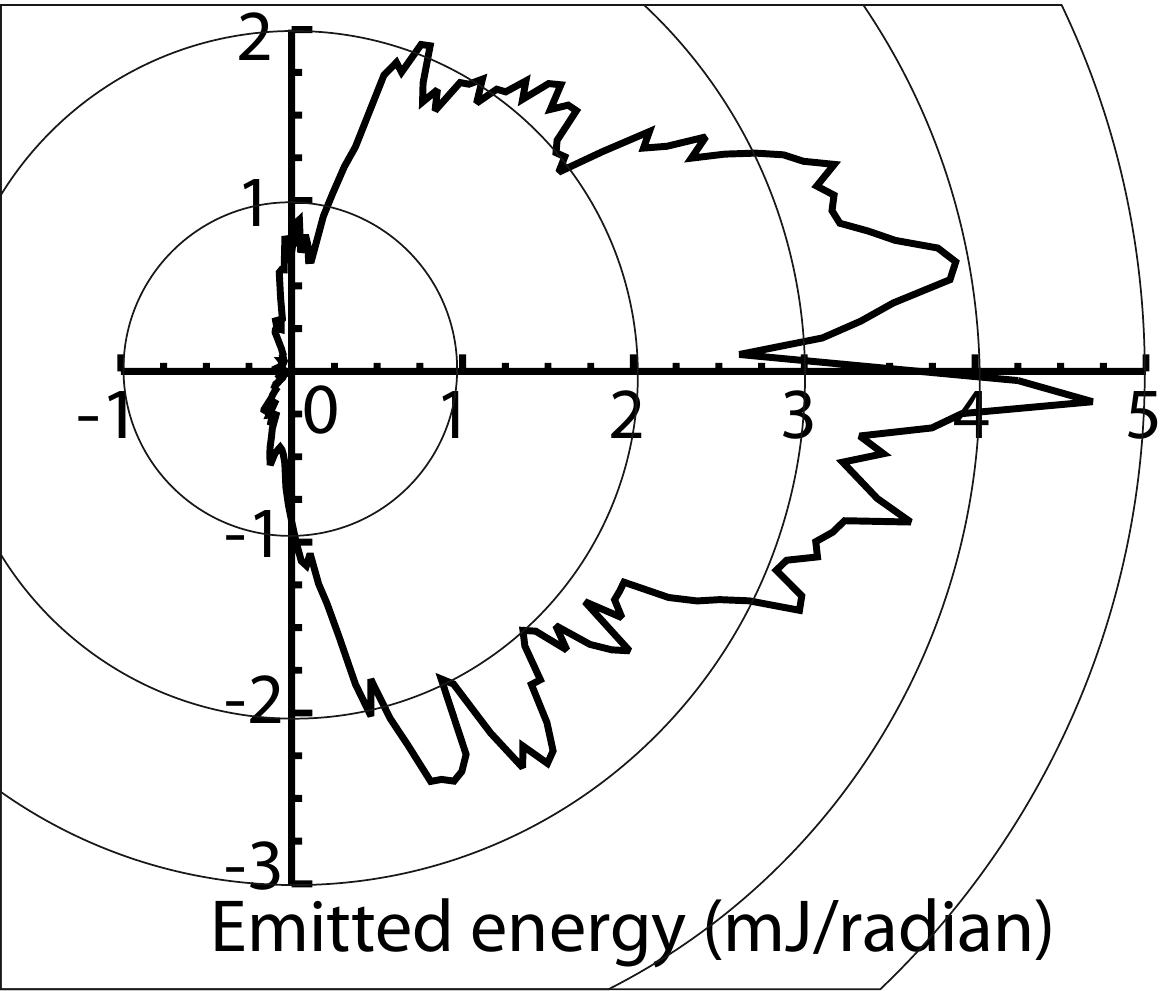}
\caption{In skin-depth emission the laser enters into a relativistically overdense target to a depth of a few skin-depths before reflecting. The interaction of the reflecting laser with the electrons accelerated forwards by the incoming laser leads to a high value of the $\eta$ parameter and gamma-ray emission. This polar plot shows the characteristic angular distribution of the emitted gamma-ray energy.}
\label{skindepth}
\end{figure}

\begin{figure} 
\includegraphics[scale=0.5]{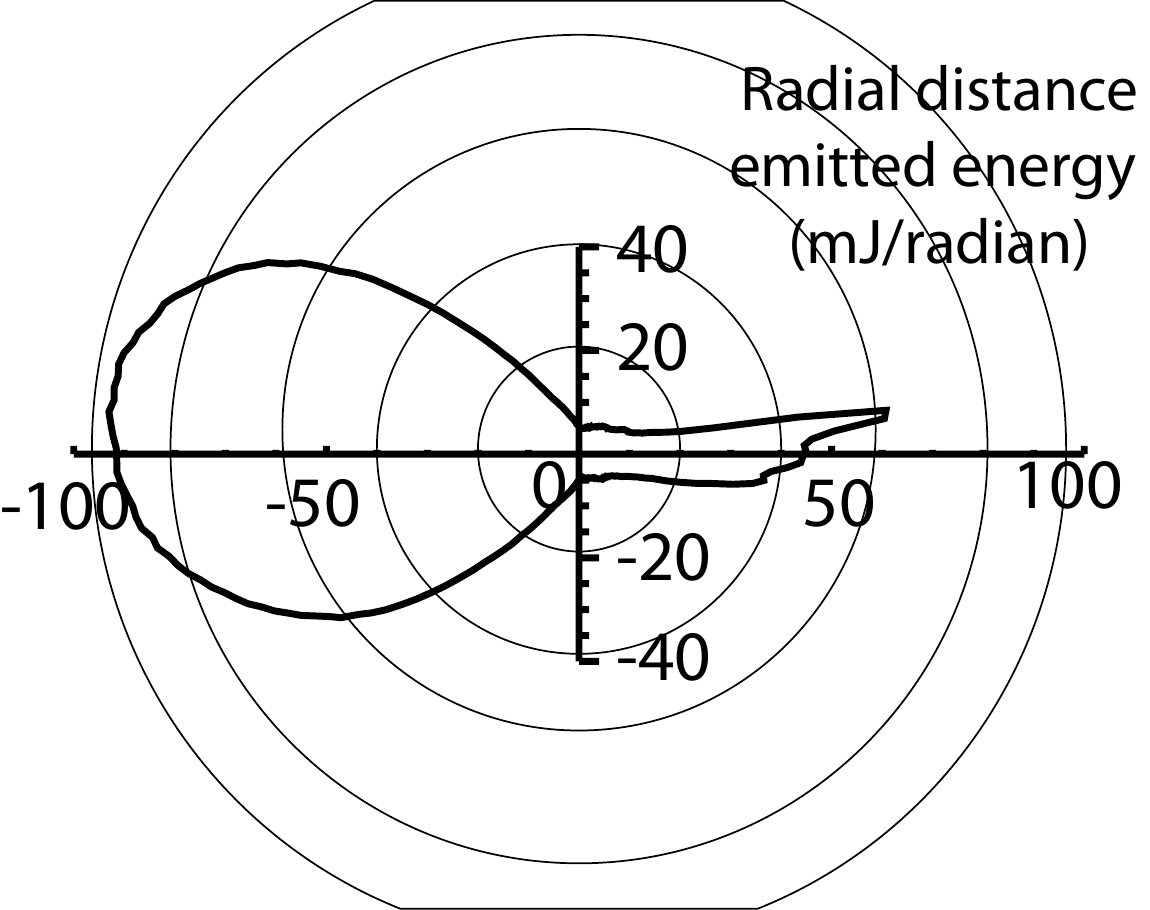}
\caption{In RESE the laser enters into a relativistically underdense target and store energy in a space charge field. This energy is released as periodic pulses of backwards propagating electrons which are in turn slowed by the radiation reaction force. This polar plot shows the angular distribution of the emitted gamma-ray energy.}
\label{rese}
\end{figure} 

\begin{figure} 
\includegraphics[scale=0.5]{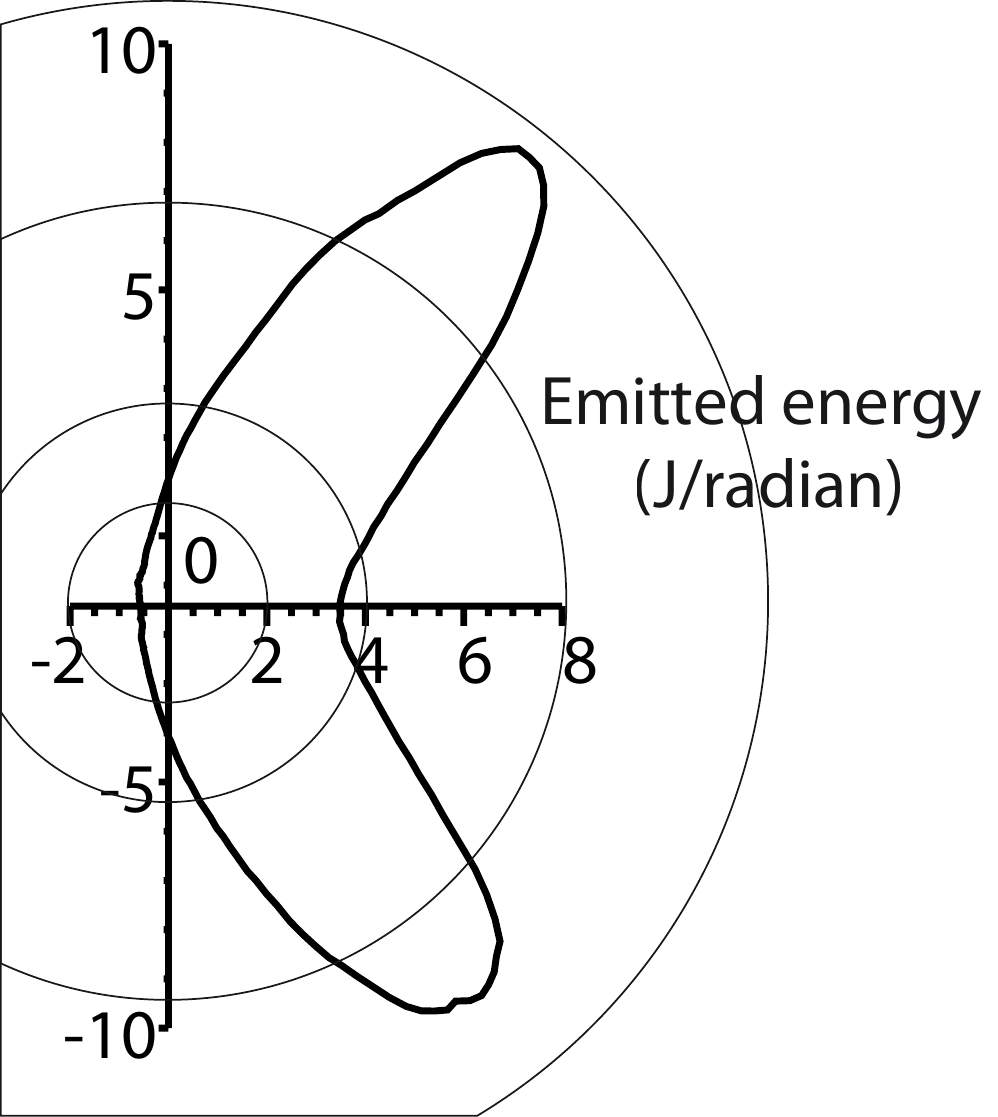}
\caption{In edgeglow transverse space charge fields cause the reintroduction of ponderomotively cleared electrons into the laser channel. This polar plot shows the angular distribution of the emitted gamma-ray energy. Emitted energy is much higher in this figure than figures \ref{skindepth} and \ref{rese} since pure edgeglow emission is only observed for lasers with intensities above $10^{24} \mathrm{W/cm^2}$.}
\label{edgeglow}
\end{figure}

In this paper gamma-ray generation by non-linear Compton scattering is modeled using the semiclassical method of Bell and Kirk \cite{Bell2008} \cite{Ridgers2013}. In this method the control parameter that determines the probability of a given electron emitting a Compton scattered photon is 
\begin{equation}
\eta = \gamma |E_\perp + \mathbf{\beta} \times c\mathbf{B}| / E_S
\end{equation} 
where $E_S = m_e^2 c^3 /q_e \hbar$ is the Schwinger electric field \cite{Schwinger1951} $E_\perp$ is the local electric field perpendicular to the motion of the electron and the other symbols have their usual meanings. The most important part of the definition of $\eta$ is that the $\mathbf{\beta} \times c\mathbf{B}$ term introduces a dependence on the angle between the electron motion and the laser's wavevector. In the limiting cases of electrons propagating either directly into or directly away from the laser $\eta$ can be written as
 \begin{equation}
 \eta = \gamma |E_\perp|/E_S (1 \pm (1-1/\gamma^2)^\frac{1}{2})
 \end{equation}
where the positive branch corresponds to an electron propagating into the laser and the negative branch an electron copropagating with the laser. Since $\gamma$ is large at intensities large enough for gamma-ray generation to be important this means that gamma-ray generation is much more important for electrons propagating into the laser than for electrons ponderomotively pushed by the laser. Intermediate $\eta$ parameters are generated by electrons propagating at an angle to the laser wavevector. While it is simple to calculate the $\eta$ parameter and hence the gamma-ray emission for a single electron, previous work \cite{Ridgers2011}, \cite{Brady2012}, \cite{Nakamura2012} has shown that global emission may be caused by different plasma physics processes \cite{Brady2013}. These processes are separated by the different ways in which they produce the electron motion which satisfies the geometrical requirement for high $\eta$.  

Skin-depth emission \cite{Ridgers2011} occurs in high density targets. RESE \cite{Brady2012} occurs in low-density targets and is the most efficient of the emission mechanisms for converting laser energy into gamma-ray energy. Edgeglow \cite{Brady2013} occurs in dense targets illuminated with highly intense lasers. These mechanisms are not mutually exclusive, but are generated by distinctly different processes which lead to electrons travelling in different directions to interact with the laser in such a way as to generate gamma-rays. Since the generated gamma-rays are emitted in a narrow cone along the direction of travel of the electron (in the model of Bell and Kirk they are assumed to be emitted in the same direction as the electron's motion) this leads to distinct angular emission profiles for each emission mechanism.

Skin-depth emission (see Ridgers et al. \cite{Ridgers2011}) is caused by the establishment of a standing wave within a few skin-depths of the target front as the laser hole-bores into the target. The backwards traveling component of the standing wave then interacts with the ponderomotively accelerated electrons generated by hole boring to produce a narrow cone of gamma-ray emission (figure \ref{skindepth}) pointing along the laser axis. The efficiency of gamma-ray generation from this mechanism is generally low both due to the reduction of the laser intensity within the skin depth and since the backwards traveling component  of the wave is of lower amplitude than the vacuum laser, reducing the $\eta$ parameter.

Reinjected electron synchrotron emission (RESE) is ultimately caused by the building up of a space charge field as electrons are ponderomotively accelerated into the target (Brady et al. \cite{Brady2012}). This space charge field eventually becomes comparable in intensity to the laser field, at which point electrons slip through the head of the laser and are accelerated back towards the laser source. This leads to the most efficient possible conversion of laser energy into gamma-ray energy for a single laser, since it gives high gamma electrons interacting with an EM wave of the full laser intensity. It produces electrons that are propagating backwards towards the laser over a broad range of angles (figure \ref{rese}).

Edgeglow is an intrinsically 2D emission mechanism caused by electrons being accelerated into the edges of the laser profile by the space charge fields set up by the transverse ponderomotive clearing of the laser bore (Brady et al. \cite{Brady2013}). It produces photons propagating in two forwards pointing lobes (figure \ref{edgeglow}) and has low efficiency since the generating electrons interact with the edges of the laser pulse which have lower intensity than the core.

These results show that there is a large range in gamma ray conversion efficiency for different laser intensities and target densities. The two emission mechanisms previously reported in Ridgers et al. \cite{Ridgers2011} and Brady et al. \cite{Brady2012} exist as two discrete regions separated rapidly by the transition between relativistic transparency and opacity. Edgeglow occurs in high density targets illuminated with high intensity lasers

\section{gamma-ray emission parameters}
 While these regimes can exist in discrete form it has also been shown \cite{Brady2013} that emission is generally a hybrid of these primitive emission mechanisms and that an important control parameter is the ratio of the target density to the relativistically corrected critical density. The parameter of real interest is the efficiency of conversion of laser energy into gamma-ray energy and, given the importance of relativistically corrected critical density as a parameter, it makes sense to consider it initially as a function of laser intensity and target density.

\subsection{Conversion efficiency}
\label{gammadens}
\begin{figure} 
\includegraphics[scale=0.5]{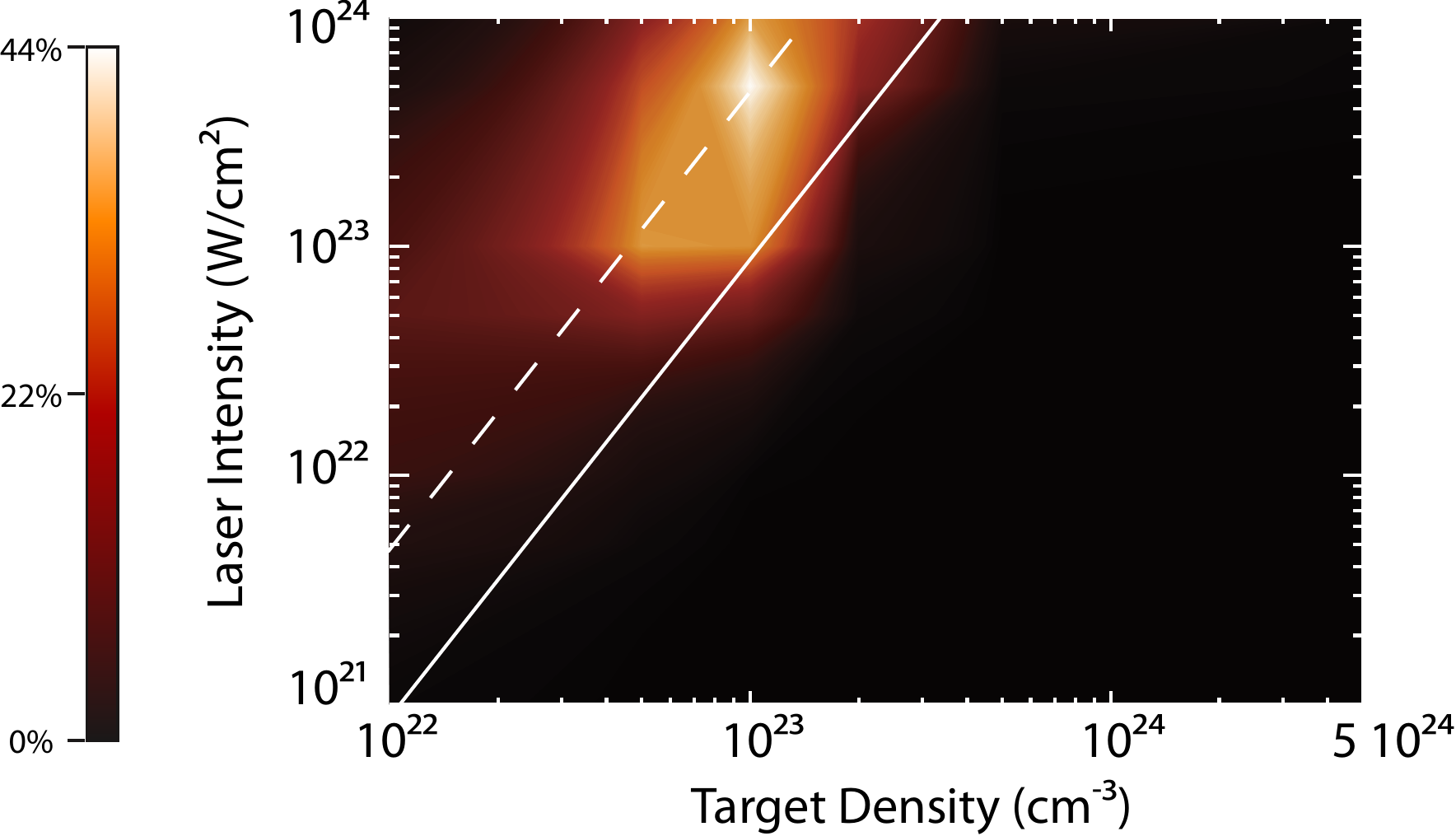}
\caption{Fraction of laser energy converted into gamma-rays as a function of target density and laser intensity for 1D simulations of an aluminium target. The solid line is the line at which the relativistically corrected plasma frequency equals the laser frequency. The dashed line is where the corrected plasma frequency is 50\% of the laser frequency. }
\label{1dresult}
\end{figure}

To compare with the previous work in Ridgers et al. \cite{Ridgers2011} and Brady et al. \cite{Brady2012} 1D particle in cell (PIC) simulations including gamma-ray generation using the EPOCH QED-PIC code were performed for 30fs long 1 micron wavelength laser pulses with intensities of between $10^{22}$ and $10^{24} \mathrm{W/cm^2}$ and an 8th order supergaussian temporal profile for a variety of fully ionized aluminium targets with an electron density of between $10^{22}$ and $5 \times 10^{24} \mathrm{cm^{-3}}$. From these simulations the fraction of laser energy converted into gamma-rays is calculated and is plotted in figure \ref{1dresult}. Overplotted as a solid white line is the line where the plasma frequency corrected using the cycle averaged electron gamma factor is equal to the laser frequency. The dashed white line is parallel to the solid line, but corresponds to the plasma frequency of 50\% of the laser frequency. There are two clear regimes on either side of the solid line: strong emission to the left of the line and much lower emission to the right. Increasing laser intensity increases both the relative and absolute production of gamma-rays. There is also a correspondence between gamma-ray production and the transition between relativistic transparency and opacity.

\begin{figure} 
\includegraphics[scale=0.5]{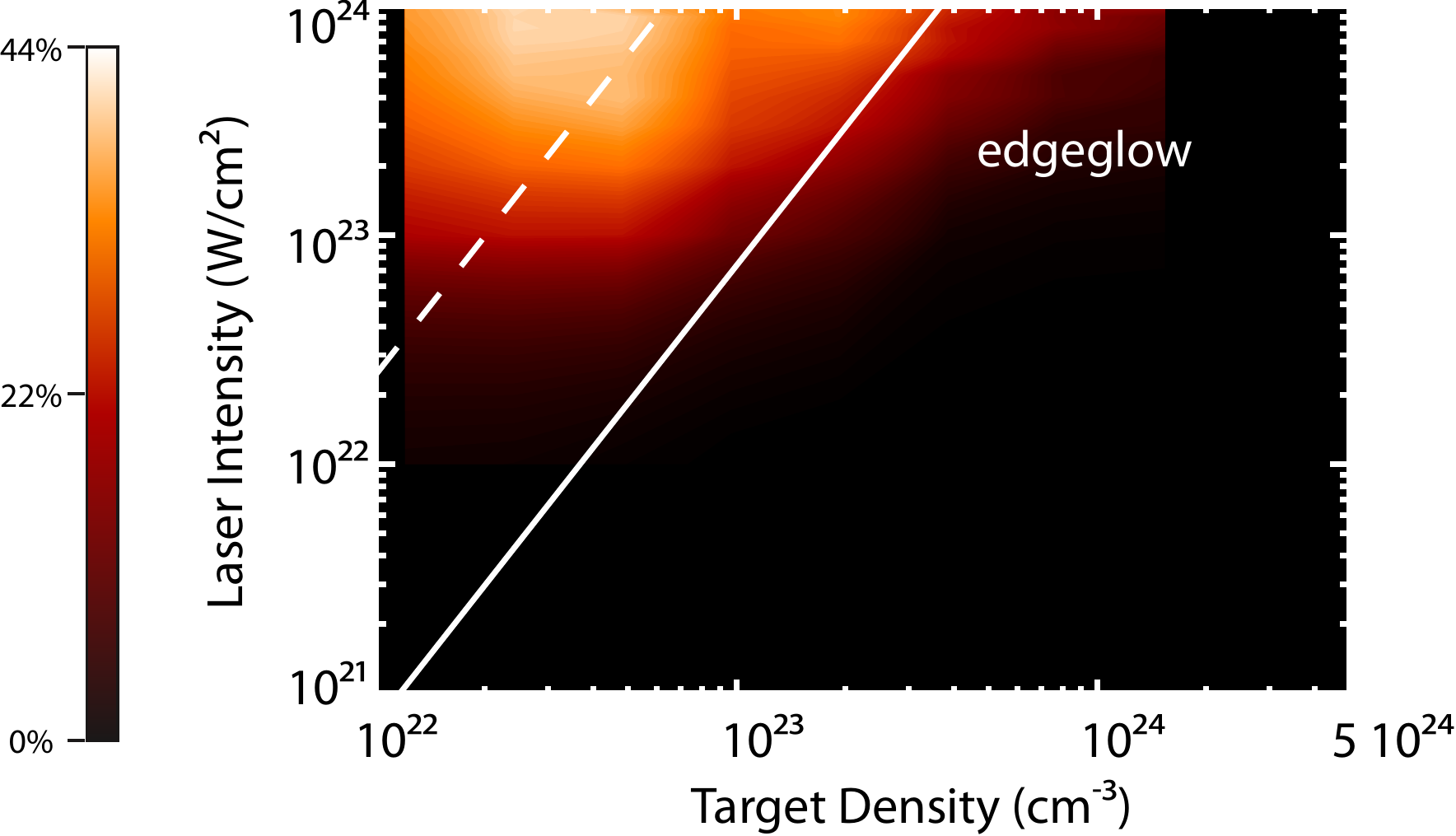}
\caption{Fraction of laser energy converted into gamma-rays as a function of target density and laser intensity for 2D simulations of an aluminium target with normal laser incidence. Lines and markings are as in figure \ref{2dresult}, except the white dashed line which is 40\% of relativistically corrected critical density.}
\label{2dresult}
\end{figure}

\begin{figure} 
\includegraphics[scale=0.5]{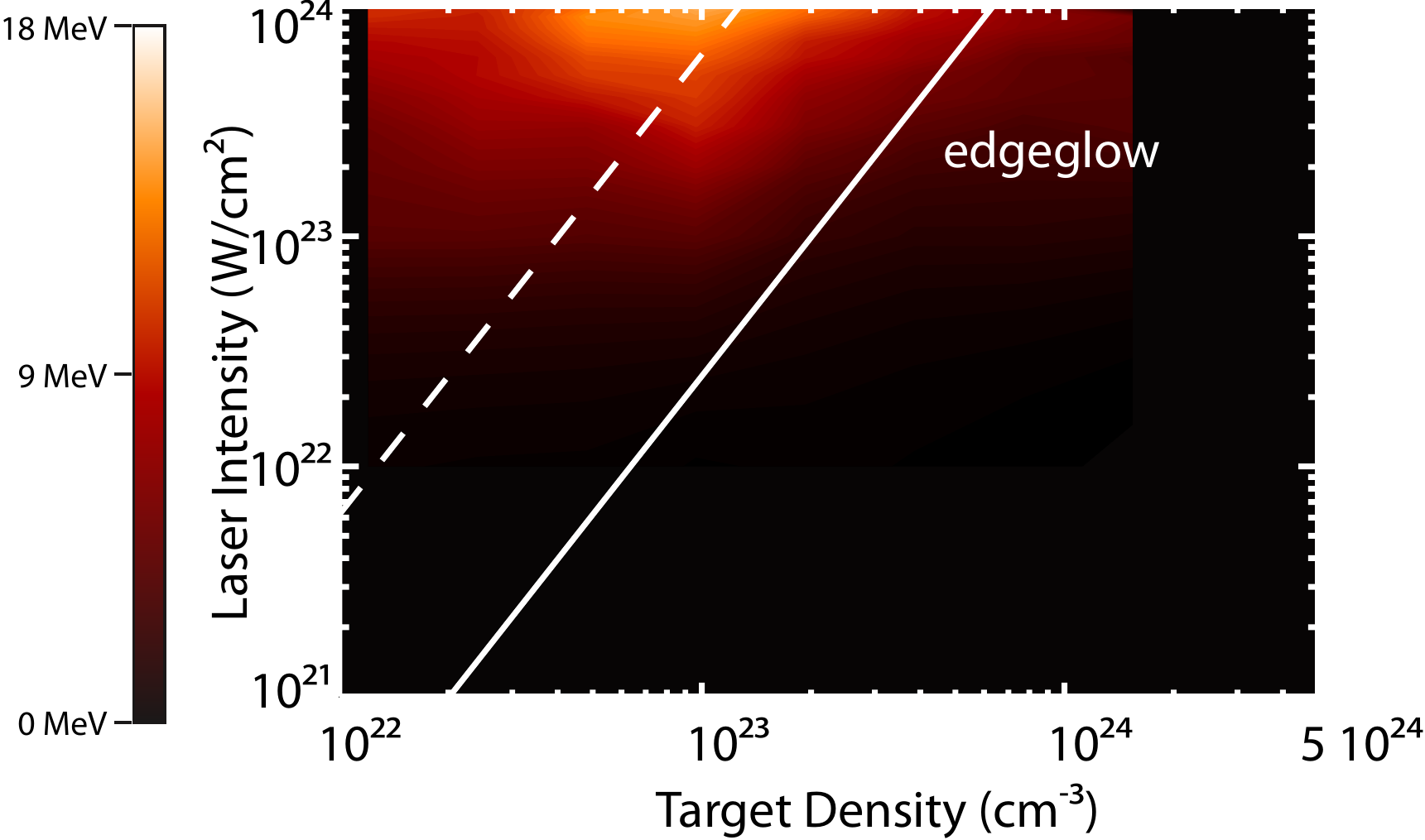}
\caption{Energy of average photon emitted as a function of target density and laser intensity for 2D simulations of an aluminium target with normal laser incidence. Lines and markings are as in figure \ref{2dresult}, except the white dashed line which is 20\% of relativistically corrected critical density.}
\label{2denergy}
\end{figure}

The pattern is similar for 2D simulations, which replicate the parameters of the 1D simulations but include a transverse spot size at focus of 1 micron (figure \ref{2dresult}) although the line fitting the maxima of emission is now at 40\% of critical rather than 50\%. The similarity between the emission efficiency contours and the line separating the underdense and overdense regimes is not as clear as in the 1D case, but it is still clear that there is some critical density (which increases as laser intensity increases) which, when exceeded, leads to a rapid decline in gamma-ray production in both 1D and 2D simulations. The only exception to this is in 2D at high intensity where weak, but higher than expected from the 1D simulations, emission is observed at high density. This is due to edgeglow emission \cite{Brady2013} and occurs in the region marked in figure \ref{2dresult}. Average photon energy (figure \ref{2denergy}) has a similar variation with density and intensity to conversion efficiency (figure \ref{2denergy}) although the density which maximizes the average photon energy is about twice that which maximizes total conversion efficiency.\\

\subsection{Angular spread of gamma-ray emission}

\begin{figure} 
\includegraphics[scale=0.5]{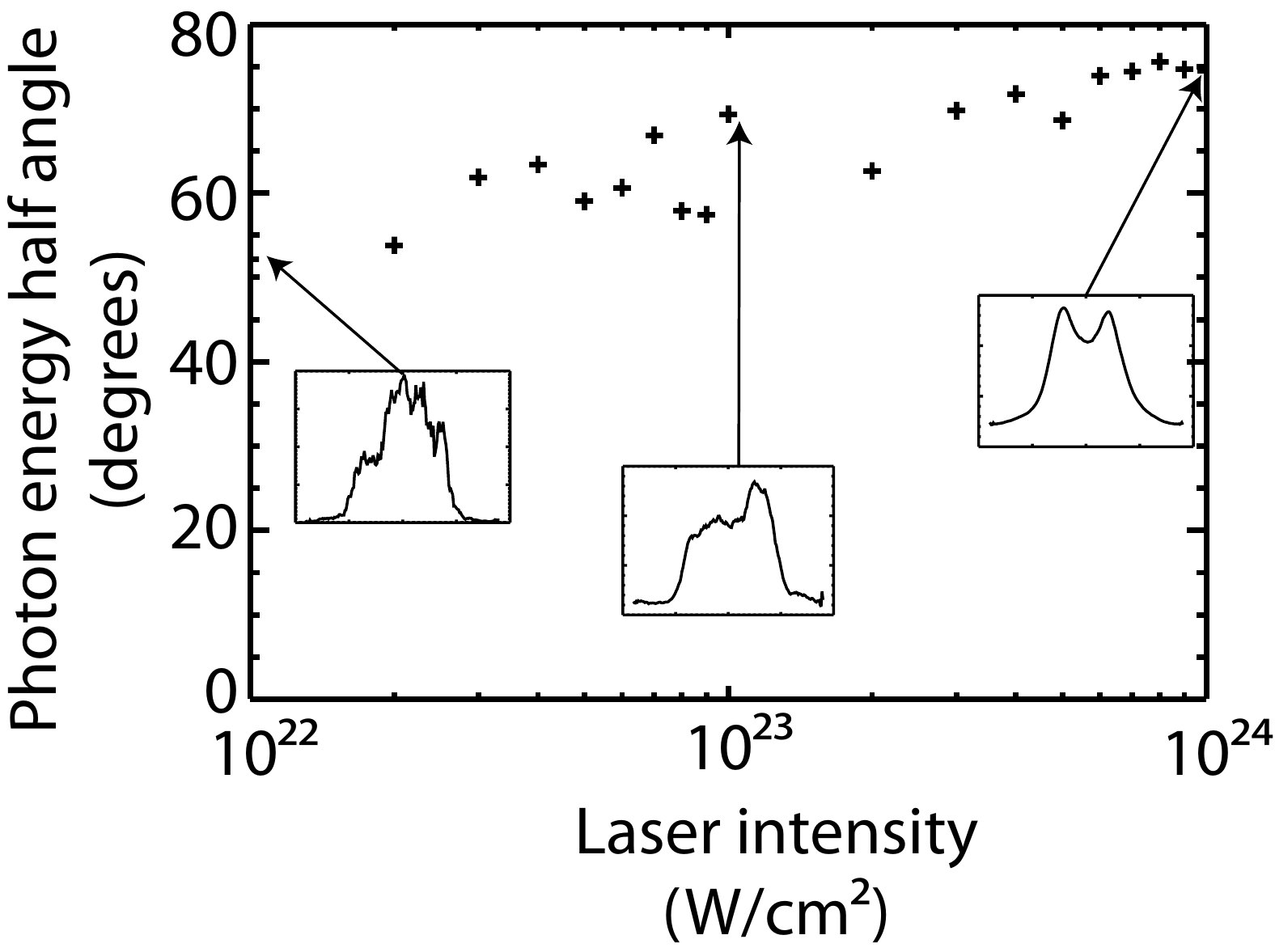}
\caption{Angular spread of all gamma-ray energy generated by the interaction of a 30fs laser pulse with a solid aluminium target for different laser intensities. A slow upward trend is visible. The inset plots show the angular distributions of emitted gamma-ray energy with the centre of the inset being directly forwards along the laser axis and the edges being $\pm180^o$ respectively. Below $5 \times 10^{23} \mathrm{W/cm^2}$ the forward peaked gamma-ray emission characteristic of skin-depth emission is seen. At $10^{24} \mathrm{W/cm^2}$ the clear double peaked structure of edgeglow is visible.}
\label{intensangle_skindepth}
\end{figure}

\begin{figure} 
\includegraphics[scale=0.5]{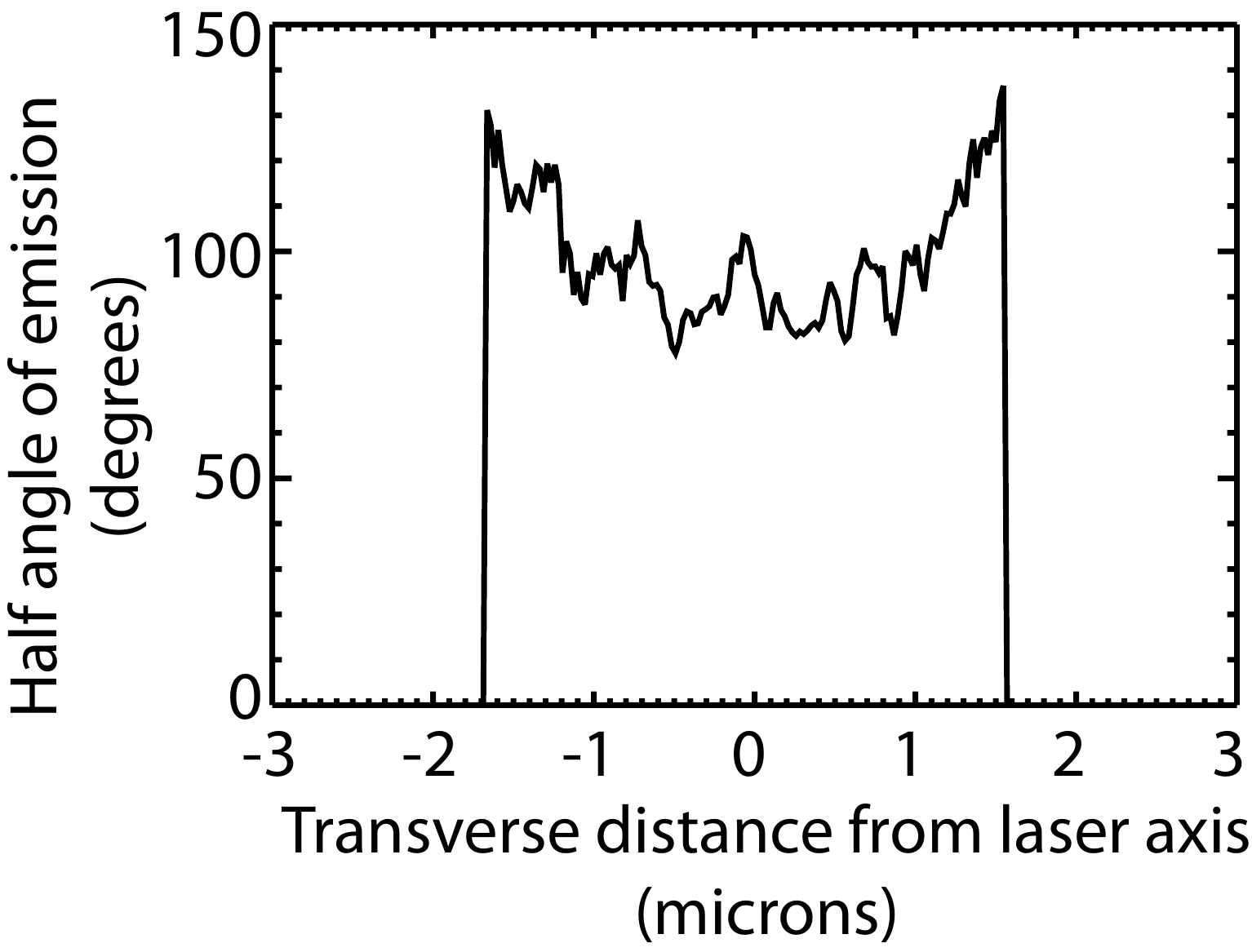}
\caption{Angular spread of photon energy varies with position of photon generation for a $5 \times 10^{22} \mathrm{W/cm^2}$ laser striking a solid aluminium target. The angular spread of emission of photons is larger for photons generated at the edge of the pulse. The angular spread in the centre is consistent with the prediction from Ridgers et al. \cite{Ridgers2011} but the overall angular spread increases with increasing photon energy due to the increasing importance of emission from the edges. To avoid errors from cells containing few photons the angle of emission is set to zero in cells that contain less than 1\% of the total photon energy.}
\label{spatialangle_skindepth}
\end{figure}

The angular width of the gamma-ray spectrum generated by the skin-depth emission model as presented in Ridgers et al. \cite{Ridgers2011} predicts that the angular distribution is $\phi_{\mathrm{sim}} = \cos^{-1} (v_{\mathrm{HB}}/c)$ where $\phi_{\mathrm{sim}}$ is the expected half angle, $v_{\mathrm{HB}}$ is the hole boring velocity from Robinson et al. \cite{Robinson2009}. The simulations in that paper provided angular widths that are broadly consistent with this prescription. However, the simulation results obtained here do not have a matching functional form as intensity increases. Instead it is found that the width of the angular emission increases slowly with increasing intensity (figure \ref{intensangle_skindepth}). This is due to skin-depth emission occuring in the edge of the laser spot. Since the hole-boring velocity is related to the laser intensity the expected angular spread of photons generated in the edges of the laser spot is greater than for those in the center of the spot. This process is not the same as edgeglow since the gamma ray generation from all transverese locations is preferentially along the laser axis, but with a wider angle of emission at the edges of the laser spot (figure \ref{spatialangle_skindepth}). As can be seen from the inset figures on figure \ref{intensangle_skindepth} edgeglow does play a role in changing the angular distribution of gamma-rays as laser intensity rises. Due to the non-linear nature of the gamma-ray emission the importance of the edges to the total photon production increases with increasing intensity leading to the increased total emission width.

Electrons in RESE emit all the way thorough the laser cycle. This means that an emitting electron can have a transverse speed anywhere from zero to the maximum quiver velocity leading to broad angular emission. The RESE mechanism described in Brady et al. \cite{Brady2012} states that the peak of the accelerating space charge field that causes the emission events is always comparable to the laser peak electric field. This means that the transverse and longitudinal velocities are always similar for all laser intensities leading to the angular width of RESE emission being independent of laser intensity. The importance of the residual skin-depth emission that leads to the forwards propagating gamma-ray component is controlled by the laser intensity. There is no dependence on target density so long as the target is not close to transparency.

Edgeglow is caused by electrons which have been ponderomotively ejected from the laser being reintroduced by the space charge field along the edge of the plasma channel. Simulations show that there is no dependence on either laser intensity or target density on the angular distribution of edgeglow emitted photons over the parameter range considered in this paper.

\subsection{Effect of target preplasma}
\begin{figure} 
\includegraphics[scale=0.5]{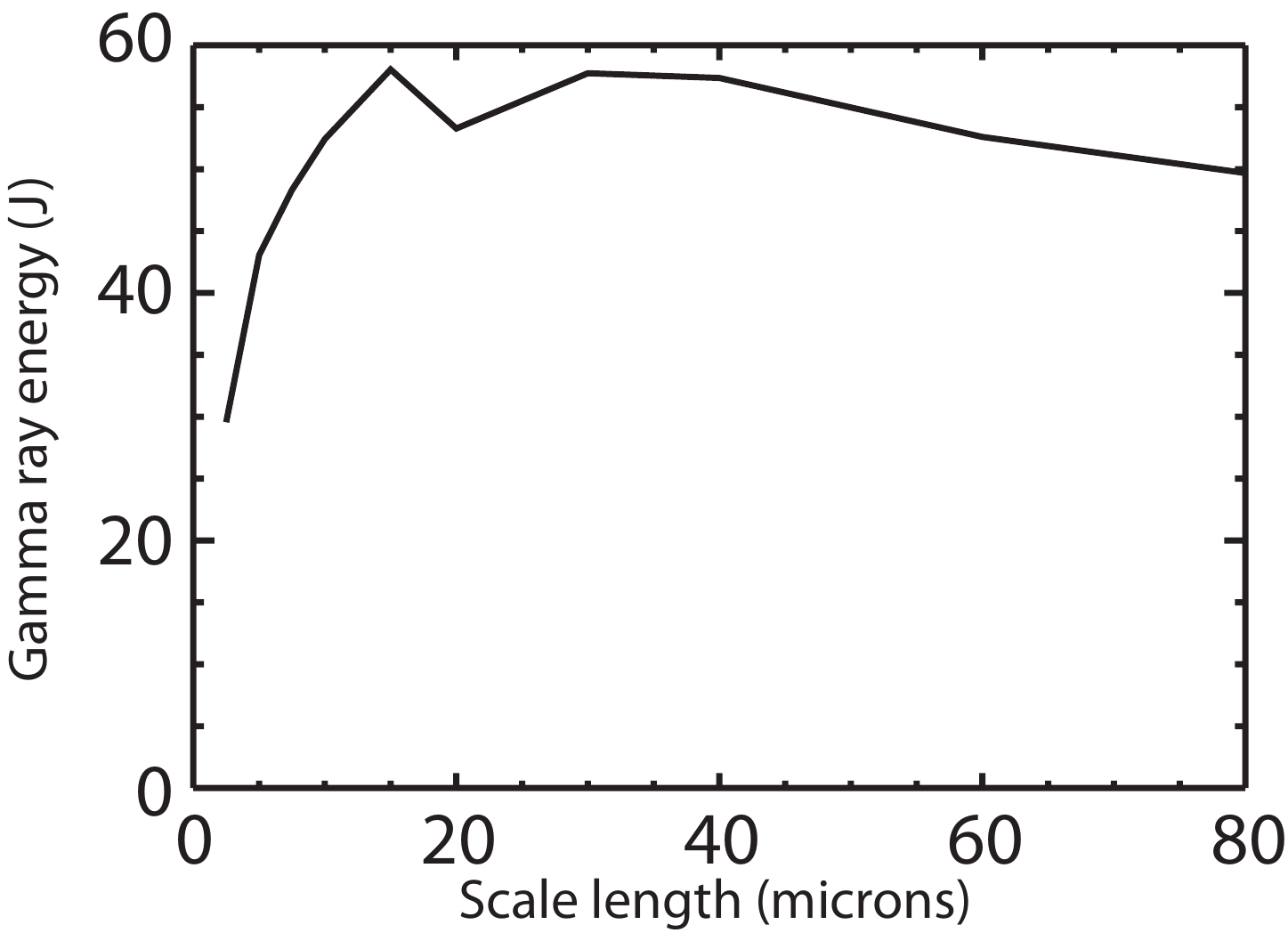}
\caption{Total photon energy produced by the illumination of a solid density aluminium target with preplasmas of different scale lengths by a $10^{23} \mathrm{W/cm^2}$ laser.  }
\label{ramp_average_energy}
\end{figure}

The work of Nakamura et al. \cite{Nakamura2012} has shown that target preplasma scale length is an important parameter for gamma-ray emission, but it is not clear from this work whether preplasmas directly affect the production of gamma-rays or whether the emission is just an averaged emission from the underdense material in the preplasma. To study this, 1D PIC simulations were performed of a $10^{23} \mathrm{W/cm^2}$ laser striking an aluminium target with various different exponential preplasma scale lengths (including no preplasma). Both the total emissivity and the electron density of material at which the emission occurred was measured. The emission of gamma-rays is maximized at a certain preplasma scale length (figure \ref{ramp_average_energy}) in a manner comparable with that observed in Nakamura et al. For short preplasma scale lengths the target is well approximated by a simple solid aluminium block and gamma-ray production is inefficient since the target is relativistically overdense. For very long scalelengths the target is well approximated by a very low-density target which is known from section \ref{gammadens} to be inefficient at producing gamma-rays since the laser accelerates electrons which do not interact with the laser to produce gamma-rays.

\begin{figure} 
\includegraphics[scale=0.5]{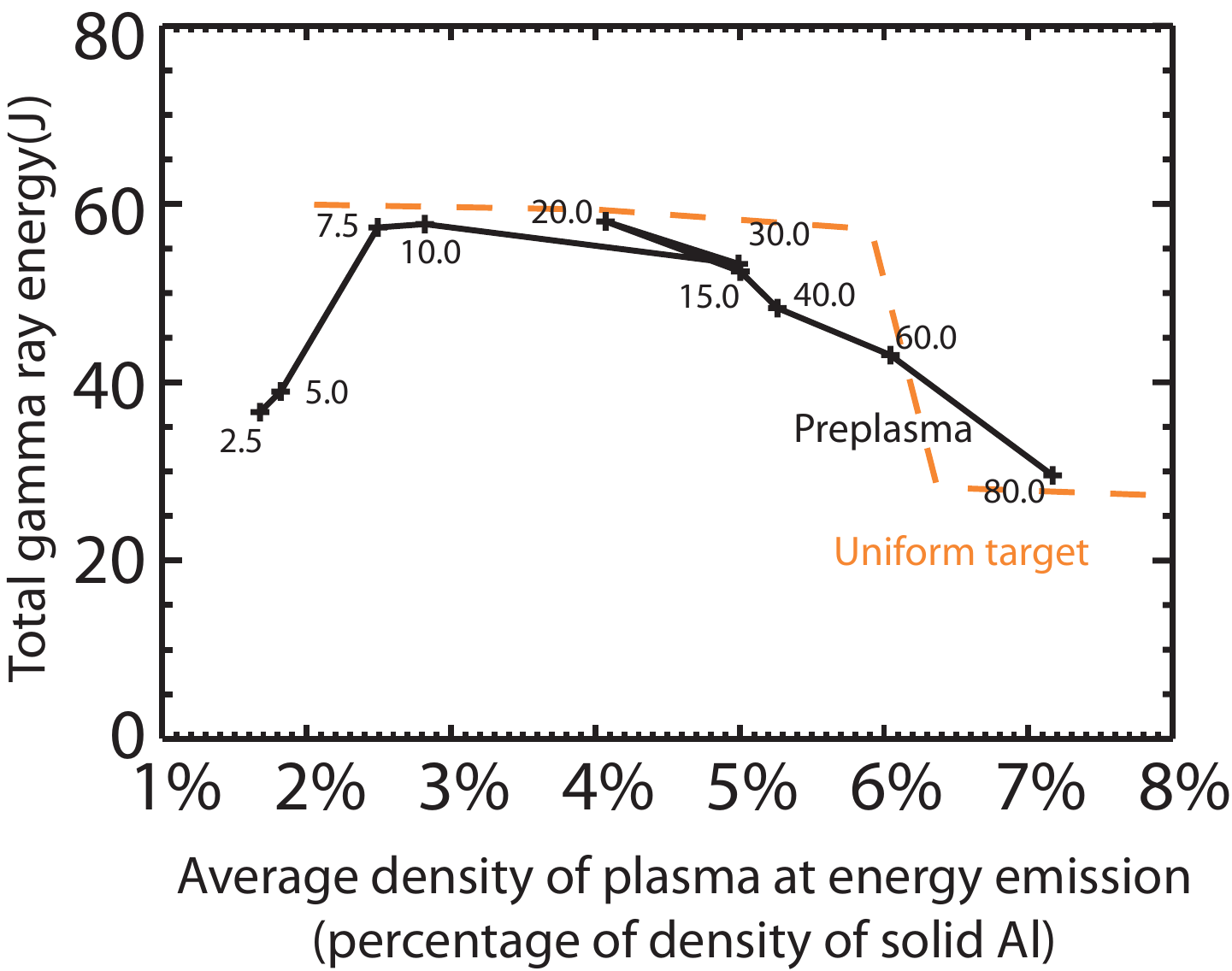}
\caption{Density of plasma in which gamma-ray energy is generated integrated over whole simulation time for a simulation of the interaction of a $10^{23} \mathrm{W/cm^2}$ laser with a target with preplasma. The numbers are the scale lengths of the target preplasma in microns. The emission density is peaked at about 3\% of the density of solid aluminium. The orange dashed line is the same figure for solid targets of different target densities}
\label{ramp_density}
\end{figure}

Analyzing the density of plasma that is emitting the gamma-rays (figure \ref{ramp_density} black line) provides an explanation of how density ramps affect the production of gamma-rays. As the pre-plasma scale length increases the average density of the emitting plasma also increases but for, scale lengths between 7.5 microns and 40 microns, even though the emitting density still increases with increasing scale length the total conversion efficiency does not. Repeating this analysis with the 1D low-density uniform targets from section \ref{gammadens} (orange dashed line on figure \ref{ramp_density}) shows that there is a similar flat region over the same range of emitting densities but with a higher emissivity. Even in these uniform targets the density of the emitting plasma is much lower than the density of the target since the emission comes from electrons reinjected into the laser rather than from the body of the target. The similarity between the lines demonstrates that the density scale length in the preplasma is not itself important in gamma-ray emission but merely provides a source of low-density plasma which leads to efficient gamma-ray generation. The existence of the flat region of maximum emission shows that there is a preferred range of densities for gamma-ray emission.

\section{Pair plasma creation}

Pair production in this parameter regime is mostly due to the non-linear Breit-Wheeler interaction since the cross section for the Trident process is orders of magnitude smaller and the Bethe-Heitler process is only important in thick targets.  Breit-Wheeler emission is simulated in EPOCH using the method detailed in Duclous et al. \cite{Duclous2010}. In this method there is a control parameter $\chi = (h\nu/(m_ec^2)) E/E_S$  for each gamma-ray photon which determines the importance of Breit-Wheeler pair production based on local conditions. Unlike $\eta$ this parameter has no geometric component meaning that if there is a high energy photon in a region of strong electric field there is a high probability of pair production.

\begin{figure} 
\includegraphics[scale=0.5]{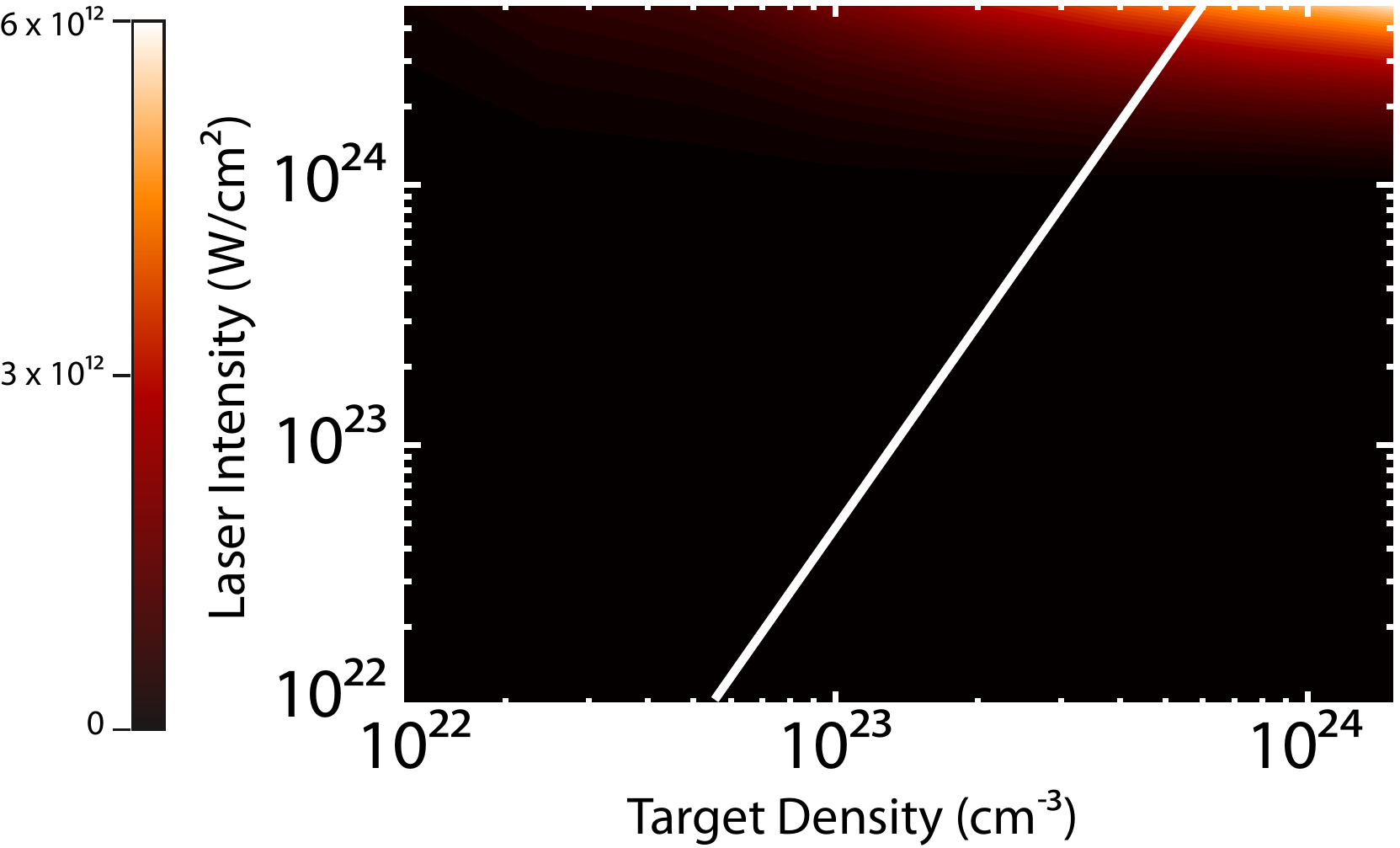}
\caption{Total number of produced pairs for different target densities and laser intensities.}
\label{posi_count}
\end{figure}

A similar phase space plot to that presented in figure \ref{2dresult} for the number of produced pairs for different laser and target parameters are shown in figure \ref{posi_count}. Pair production is maximized for high laser intensities and high target densities and there is no association between the relativistically corrected critical density and the total number of pairs created.  

Pair production is most efficient in the region where photon production is primarily due to edgeglow emission. This is because edgeglow emits photons into two forward pointed lobes configured so that the upwards pointing lobe is generated by the lower cut edge and the downwards pointed lobe is generated by the upper edge; the photons are generated so that they are travelling towards the laser core and are guaranteed to interact with the highest intensity part of the laser. This means that edgeglow is a perfect emission mechanism for pair production: produced photons tend to be high energy and they are produced propagating in a direction that means they will interact with the high intensity region of the laser. 

\section{Relativistic transparency and longitudinal electron dynamics in the ultra-relativistic regime}
\begin{figure*} 
\includegraphics[scale=0.5]{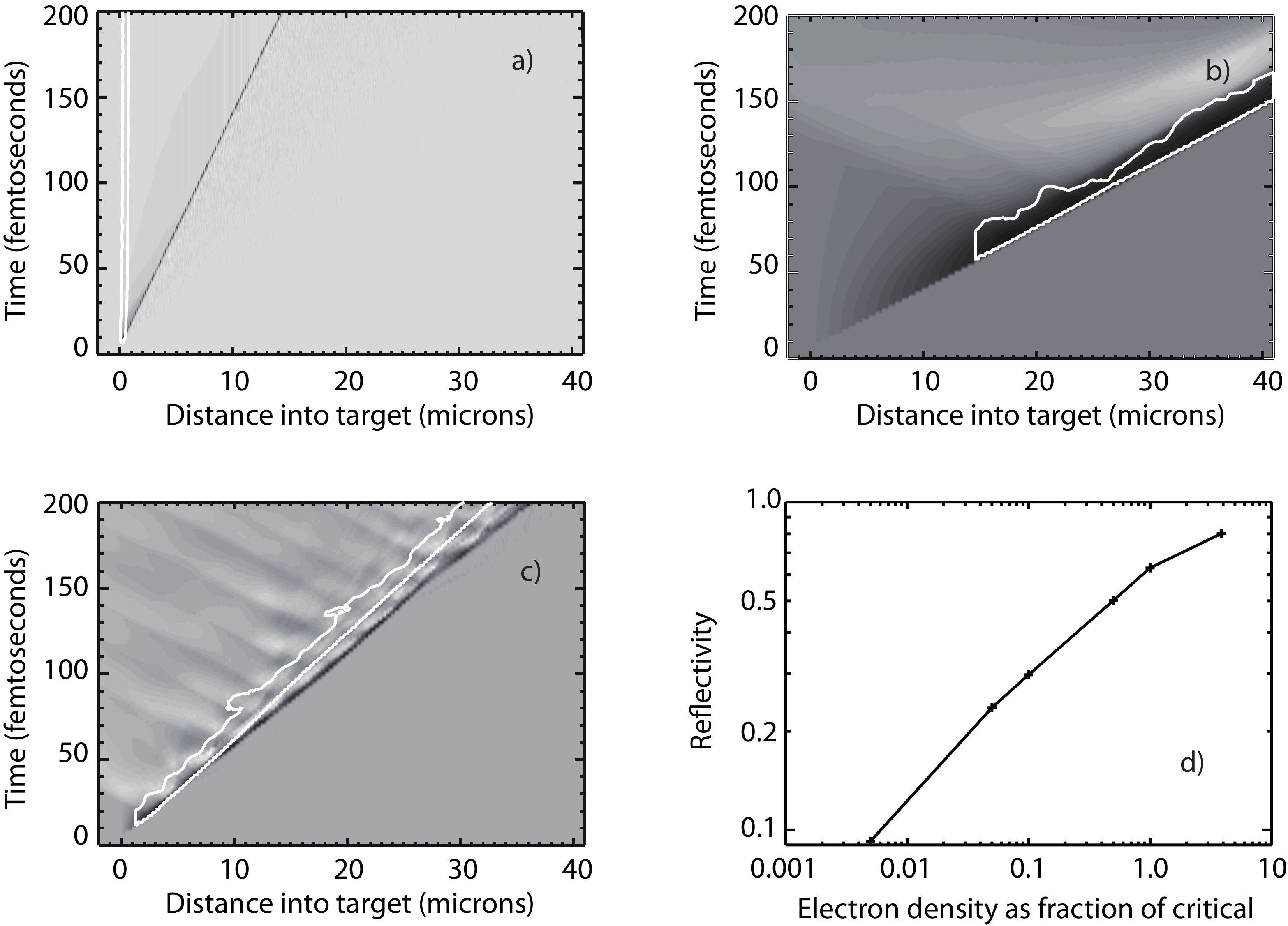}
\caption{Parallel electric field in time for three different target densities illuminated with a $2 \times 10^{23} \mathrm{W/cm^2}$ circularly polarized laser both when the ions are infinitely massive and for aluminium ions. The location of the front of the laser is shown as the black line on the coloured contours. The solid colours are for aluminium and the thick white contour line are for immobile ions. a) is for a target of relativistically corrected critical density, b) is for a target of 1\% critical density and c) for 10\% of critical. It is clear that as the target density decreases the influence of the ions on the propagation of the laser into the target decreases. d) shows the increase in laser intensity reflectivity for higher target densities. This increase is monotonic and does not show any sudden jump at the point where a pure electron plasma no longer shows a propagating front.}
\label{figure1}
\end{figure*}

There is no doubt that there is an association between the transition between relativistic transparency and relativistic opacity and the transition between high efficiency RESE and low efficiency skin depth emission. At the simplest level relativistic transparency or opacity is simply the answer to the question "does the laser produce a transverse current in the plasma such that the laser is reflected?", but the difference between RESE and skin depth emission is in the longitudinal electron dynamics, not the transverse. This then asks the question of whether the transition between the two is really associated with relativistic transparency or some other, longitudinal, process that has similar scaling with laser intensity and target density or even whether the two processes are mixed. The last view is supported by work such as that of Cattani et al. \cite{Cattani2000} where it is shown that effects like ponderomotive pileup make transparency different in the high powered laser regime. 

Despite these semantic points, the existence of two distinct regimes separated by the ratio of the relativistically corrected plasma frequency to the laser frequency is clear and detectable by identifying the importance of ion motion in the plasma dynamics. In both regimes the interaction of the laser with the plasma leads to the production of a propagating front of electron density which sets up a parallel electric field. Using the PIC code EPOCH1D \cite{Brady2011} we performed multiple simulations of a $10^{23}\mathrm{W/cm^2}$ circularly polarized laser interacting with targets of different densities with both mobile and immobile ions and tracked the location of the leading edge of the parallel electric field structure. This leading edge represents a measure of the location of the laser in the target. Where the target density is above that conventionally defined as relativistically critical there is a clear difference between the simulations with mobile and immobile ions (figure \ref{figure1} a). With mobile ions  the motion of the front shows conventional hole-boring with a speed which matches within 2\% the hole boring velocity from Robinson et al. \cite{Robinson2009}. With immobile ions the electrons are briefly pushed forwards early in the simulation, setting up a charge separation field, and are then held stably at the point where the force due to this field is equal in magnitude to the ponderomotive push of the laser. 

For simulations where the target density is much lower than relativistically corrected critical (figure \ref{figure1} b) a completely different situation is observed: the presence or absence of mobile ions has no effect on the propagation of the front through the target, and that front propagates at a different speed than would be expected from the Robinson et al. hole boring velocity. This is a demonstration of the decoupling of ions and electrons observed as a critical part of the RESE mechanism in Brady et al. \cite{Brady2012} and is the same mechanism as the change from conventional to incomplete hole boring in Weng \cite{Weng2012_1}. Between these two extremes is a region (figure \ref{figure1} c) where ions are important to the motion of the electrostatic front but even in the absence of ion motion the front still propagates. The transition between the two regimes is clearly not sudden since the reflectivity, which can be used a proxy for front velocity, increases smoothly as a function of target density (figure \ref{figure1} d)). The backwards propagating streaks visible in the parallel electric field in figure \ref{figure1} d are the discrete breakdown events that are associated with RESE.

To explain the behaviour it is necessary to consider the microscale physics of the head of the laser pulse. The electrons start to quiver due to the laser E field and then are ponderomotively accelerated parallel to the laser axis. This in turn sets up a space-charge field that resists the motion of the electrons and this field, in turn, moves the ions which cancel enough of the space charge field to allow the entire front structure to propagate into the target. If the ions are held immobile then this cancellation will not occur and a naive expectation would be that a force balance would be achieved between the space charge field and the $\mathbf{v} \times \mathbf{B}$ force leading to exactly the situation that is observed in the simulations of overdense targets. What requires explanation is what occurs in the low density targets that prevents this force cancellation from occurring.

\begin{figure} 
\includegraphics[scale=0.5]{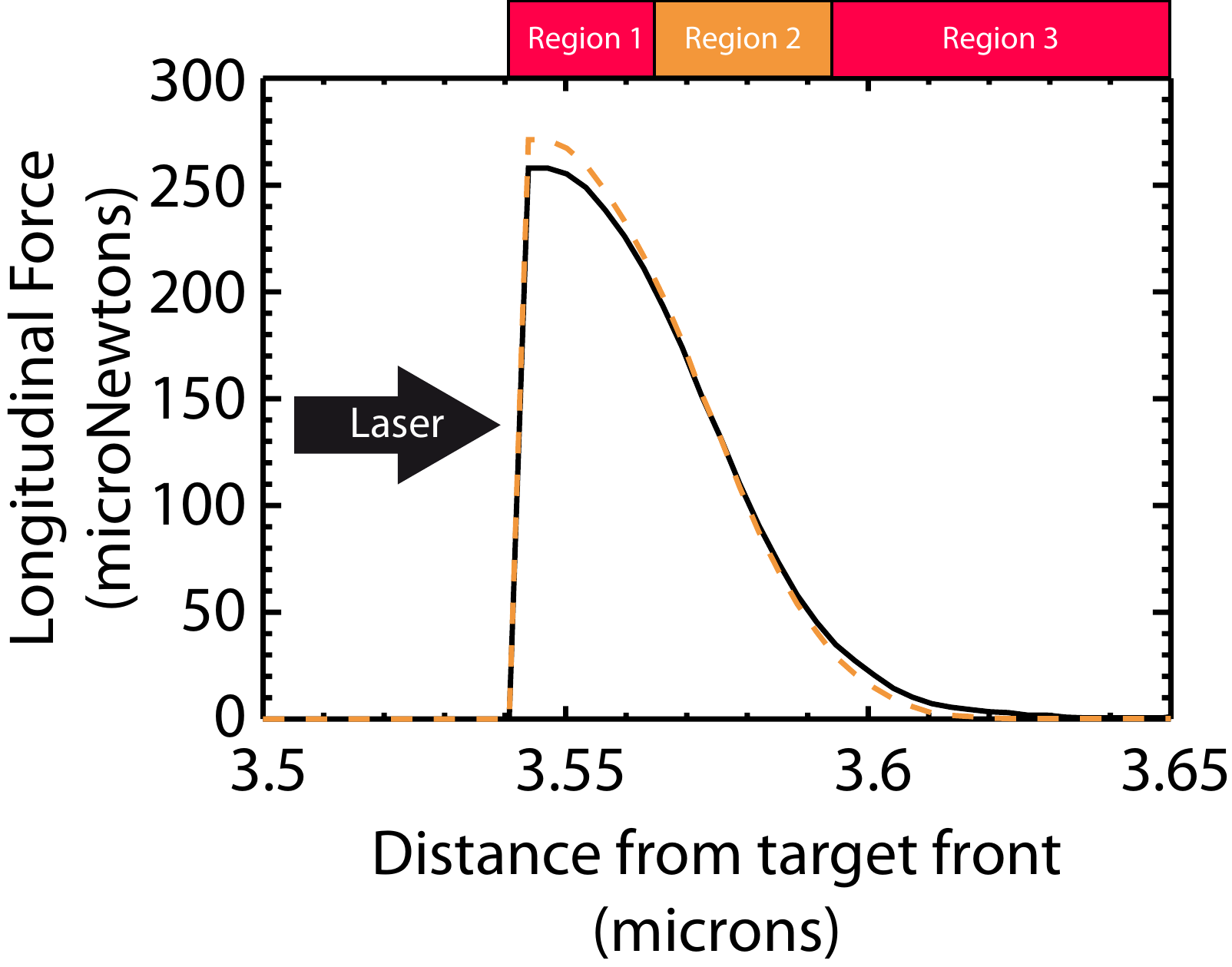}
\caption{The magnitude of the longitudnal forces on a single electron at the head of a $10^{23} \mathrm{W/cm^2}$ laser interacting with a target of electron density $7.7 \times 10^{23} \mathrm{cm^-3}$ (relativistically overdense, comparable to solid aluminium) for a simulation with immobile ions. The solid black line is the magnitude of the space-charge force due to the electron longitudinal motion. The orange dashed line is the $\mathbf{v} \times \mathbf{B}$ force due to the electron quiver motion. The $\mathbf{v} \times \mathbf{B}$ is larger than the space charge force in region1, and the space charge force is larger than the $\mathbf{v} \times \mathbf{B}$ force in region 3.}
\label{overdense_force}
\end{figure}

\begin{figure} 
\includegraphics[scale=0.5]{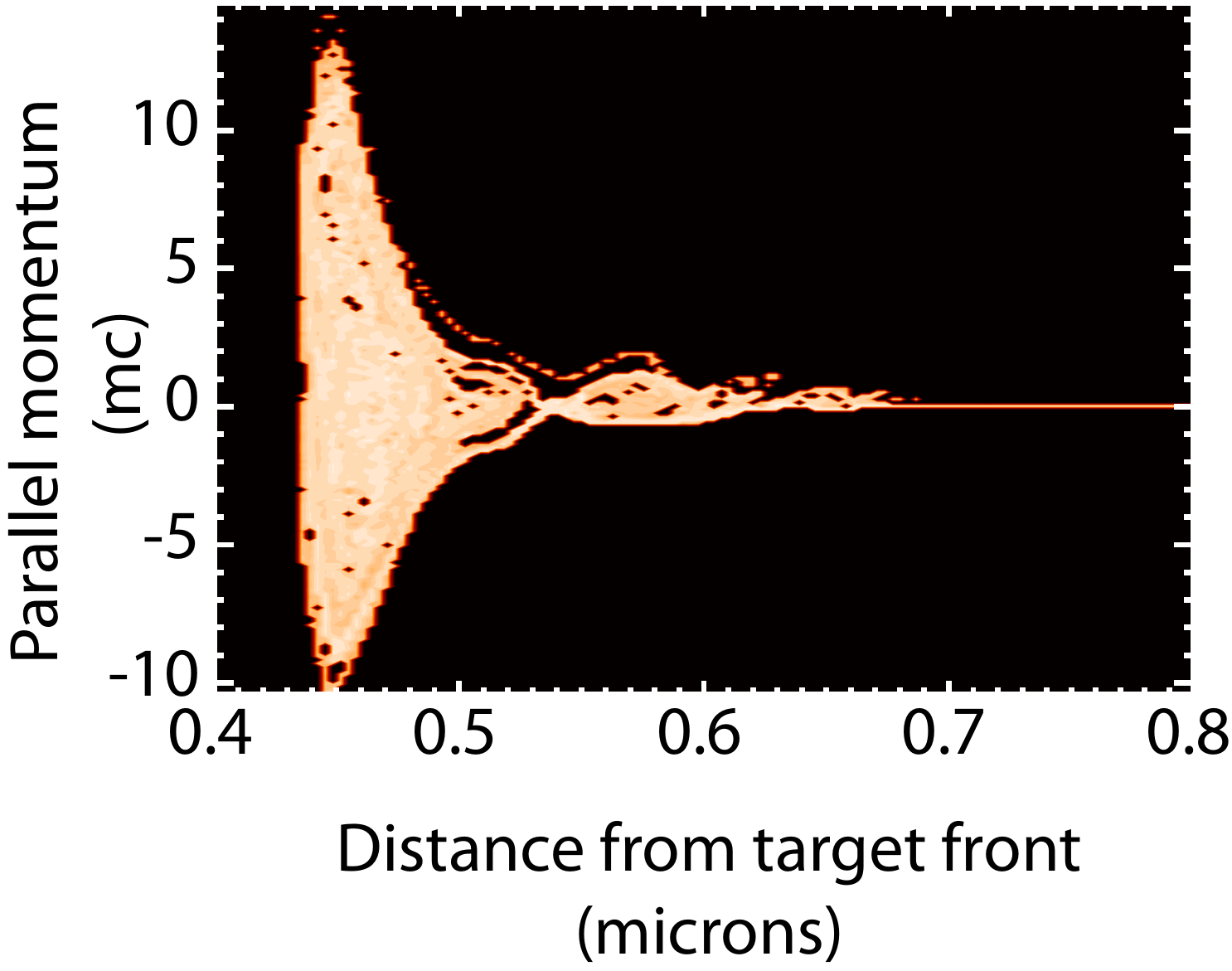}
\caption{Electron $x - p_x$ phase space plot for a $10^{23} \mathrm{W/cm^2}$ laser interacting with a target of electron density $7.7 \times 10^{23} \mathrm{cm^-3}$. Due to the opposing forces from the target front and inside the foot electrons are swept into an orbiting structure in phase space.}
\label{overdense_phase}
\end{figure}

Studying the structure of the space-charge and $\mathbf{v} \times \mathbf{B}$ forces on electrons at the laser head for different target densities shows that there is a difference between the high density and low density cases. In the overdense case there are three distinct regions (figure \ref{overdense_force}). At the front of the target, in region 1, the laser causes an excess $\mathbf{v} \times \mathbf{B}$ force which pushes the electrons in the target forwards. This is the initial mechanism which sets up the space charge field. Inside the target there is region 2 where the space-charge field and the laser $\mathbf{v} \times \mathbf{B}$ forces balance. Deeper into the target, in region 3, there's a net force due to the space charge field pushing electron towards the laser from inside the target. The fact that the lengthscales for the space-charge and $\mathbf{v} \times \mathbf{B}$ forces do not have to be equal leads to the phase space structure in figure \ref{overdense_phase}. Due to the structure of the longitudinal force shown in figure \ref{overdense_force} electrons are not just passively accelerated forwards by the laser $\mathbf{v} \times \mathbf{B}$ force but are first accelerated backwards by the space-charge field in region 3 before then encountering the laser. These backwards accelerated electrons produce the backwards propagating (lower) lobe in the phase space structure in figure \ref{overdense_phase}. During the initial "sweeping-up" phase the low strength of the space charge field compared to the laser field means that electrons travelling forwards have higher momentum than those travelling backwards and the phase space is asymmetric with forwards propagating electrons having much higher momenta than backwards propagating. As the laser head continues to propagate into the target the space charge field increases in magnitude leading to electrons being accelerated backwards to ever higher momenta and symmetrizing the phase space structure. Eventually the two lobes become equal, leading to a current free structure at the head of the laser pulse. This structure is stabilized by an excess forwards force at the target front and an excess negative at the front of the plasma structure. If there is an increase in the $\mathbf{v} \times \mathbf{B}$ force then it generates an increased space charge field which in turn restores the equilibrium and vice-versa. Since there is no net current and no net force then the plasma structure comes to a halt and propagation of the laser into the target stops. With mobile ions the $\mathbf{v} \times \mathbf{B}$ force is irrelevant and the ions are accelerated into the target by the space charge field leading to hole-boring.

\begin{figure} 
\includegraphics[scale=0.5]{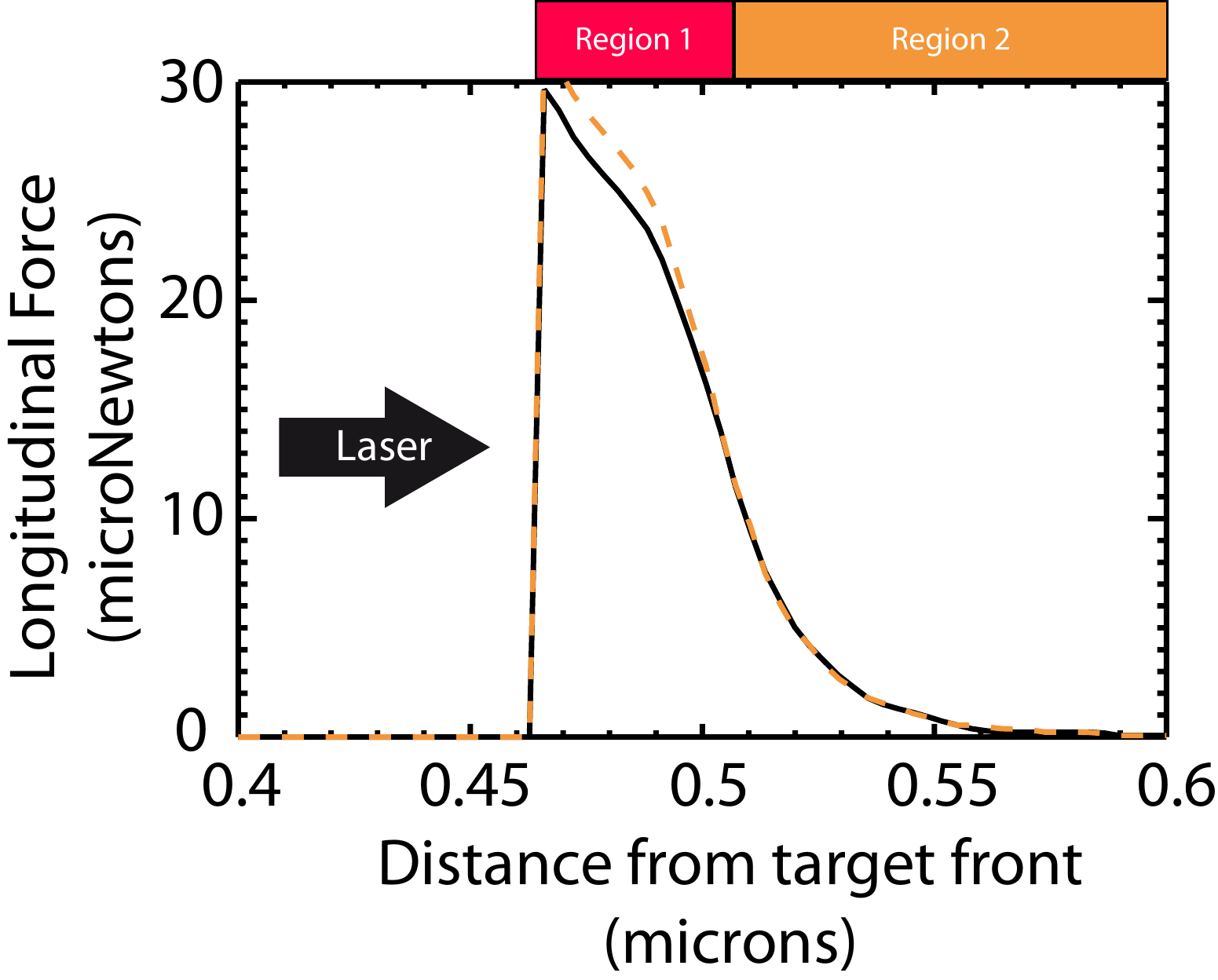}
\caption{The magnitude of the average longitudnal forces on a single electron at the head of a $10^{23} \mathrm{W/cm^2}$ laser interacting with a target of electron density $7.7 \times 10^{22} \mathrm{cm^-3}$ (relativistically underdense) for a simulation with immobile ions. The solid black line is the magnitude of the space-charge force due to the electron longitudinal motion. The orange dashed line is the $\mathbf{v} \times \mathbf{B}$ force due to the electron quiver motion. The $\mathbf{v} \times \mathbf{B}$ is larger than the space charge force in region 1, but otherwise there is near exact force balance at all points in the target.}
\label{underdense_force}
\end{figure}

\begin{figure} 
\includegraphics[scale=0.5]{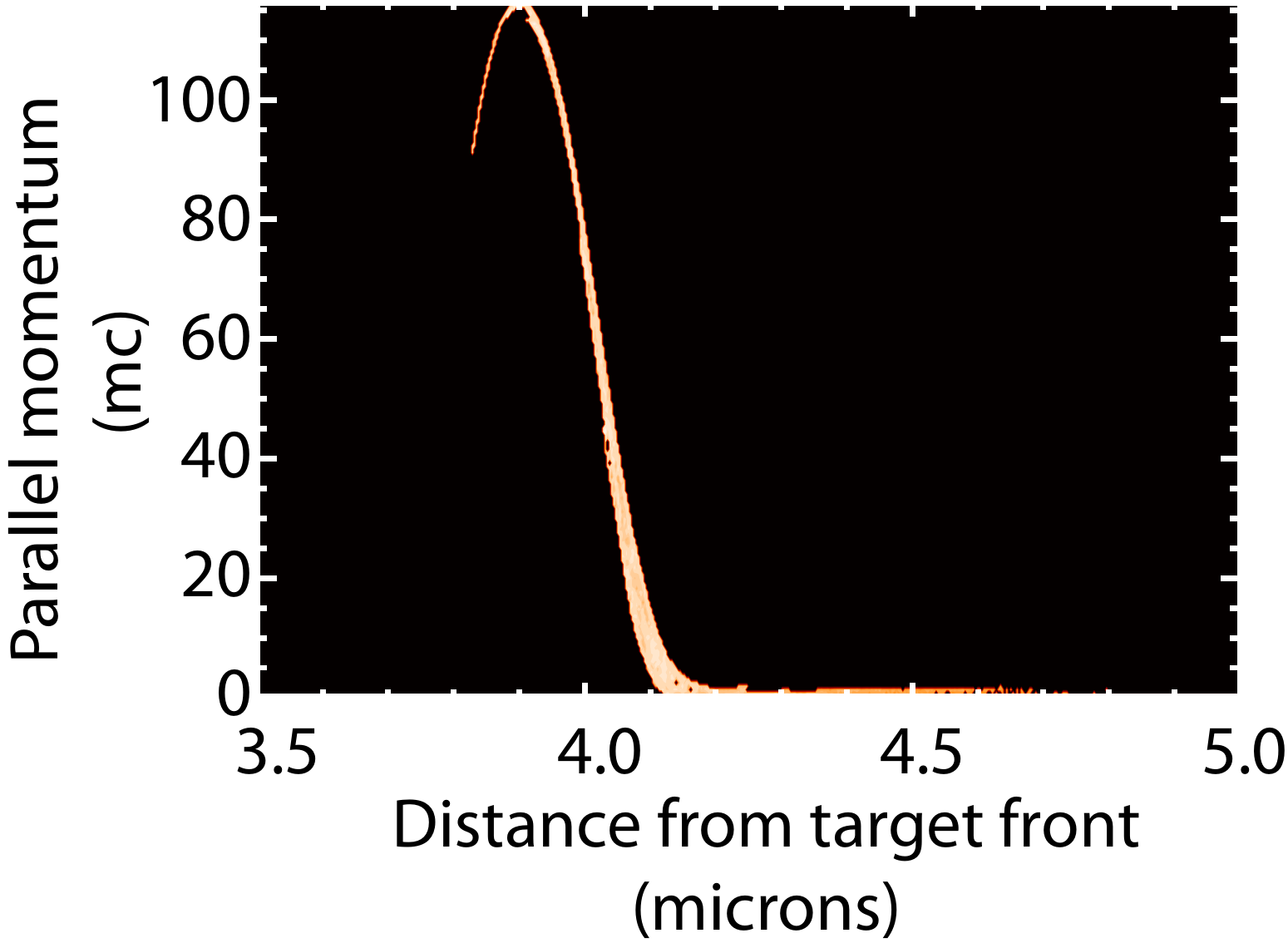}
\caption{Electron $x - p_x$ phase space plot for a $10^{23} \mathrm{W/cm^2}$ laser interacting with a target of electron density $7.7 \times 10^{22} \mathrm{cm^-3}$. Since the only net force is the forwards directed $\mathbf{v} \times \mathbf{B}$ force electrons are pushed into an asymmetrical structure with electrons either travelling forwards or undisturbed.}
\label{underdense_phase}
\end{figure}

In the underdense case the force structure at the laser head is different (figure \ref{underdense_force}) with just two distinct regions. Region 1 still exists but region 2 now extends all the way into the target and region 3 is non-existent. This in turn leads to a phase space structure which is completely different (figure \ref{underdense_phase}). In the underdense case the phase space structure is asymmetric; electrons are either accelerated forwards by the laser or undisturbed, with no electron being accelerated backwards towards the laser until a RESE breakdown event occurs.  This means that the space charge force continues to build up, eventually becoming of equal magnitude to the $\mathbf{v} \times \mathbf{B}$ force at all points. At this time, new electrons encountered by the laser experience no net forward force and slip through the laser head (this is the incomplete hole-boring behaviour of Weng et al. \cite{Weng2012_1}). Behind the laser head the force structure gives them a net backwards force which accelerates the electrons rapidly towards the laser, both canceling the space-charge field and allowing the laser to propagate into the target, but also leading to the observed RESE emission from these backwards propagating electrons. Eventually a combination of the laser $\mathbf{v} \times \mathbf{B}$ force and the loss of momentum due to the emission of gamma ray photons stops these backwards propagating electrons and the sweeping up process begins again, leading to the observed periodic emission of RESE.

\section{Conclusions}
The aim of this paper is to investigate the parameters which control the production of both gamma-rays and electron-positron pairs in linearly polarized laser solid interactions. To this end several simulations were performed evaluating the efficiency of these processes for a range of parameters. It was found that the most important parameters were target density and laser intensity and both were varied over physically relevant scales for next generation laser solid interactions. It was found that gamma-ray generation was connected with relativistic transparency with maximum gamma-ray generation occurring at some fraction of the relativistically corrected critical density for all laser intensities. This association with relativistic transparency is explained in terms of microphysical behaviour at the head of the laser pulse as it propagates into the target, which splits the phase space into two sections. In low-density transparent plasmas the backwards propagating electrons responsible for RESE emission are observed \cite{Brady2012}. In high density plasmas the accelerated electrons are confined into a thin skin layer at the laser head leading to the less efficient skin-depth emission from Ridgers et al. \cite{Ridgers2011}.  A similar association with relativistic transparency is also spotted for the efficiency of ion acceleration. Preplasmas and density ramps are shown to be important only in so far as they provide a source of low-density plasma allowing strong gamma-ray production by RESE emission.

Breit-Wheeler pair production is maximized at high target density and high laser intensity. This is explained in terms of the requirement that electrons must reach high enough relativistic gamma factors to allow significant production of gamma-rays with high enough energy to produce pairs. In regions where gamma-ray production is efficient the reduction in the average electron kinetic energy due to the radiation reaction force turns out to be more important than the increase in the number of photons and results in a decrease in the number of photons produced with sufficient energy to give a high probability of pair production. Other parameters such as pulse length, pulse temporal envelope and pulse transverse profile were simulated and found to have little or no effect on conversion efficiency of laser energy into gamma-rays or electron-positron pairs.

Next generation lasers have a wide possible range of applications in science, medicine and industry, but these applications take advantage of different properties of the lasers. The results in this paper show that by changing the laser intensity and the target density it is possible to move between regimes of efficient gamma ray production, efficient pair production and more or less conventional plasma physics where additional QED-plasma processes are not energetically important. The results in this paper produce a possible starting point for developing these different applications.

\section*{Acknowledgement}
The EPOCH code was developed as part of the UK EPSRC funded projects EP/G054950/1 and EP/G056165/1

\section*{References}
\providecommand{\noopsort}[1]{}\providecommand{\singleletter}[1]{#1}%

\end{document}